\documentclass[aps, twocolumn, eprint]{revtex4-1}
\usepackage{amsfonts}
\usepackage{amssymb}
\usepackage{amsmath}
\usepackage{graphics}
\usepackage{graphicx}
\usepackage{subfigure}
\usepackage{dcolumn}

\usepackage{bm}
\usepackage{color}
\usepackage{epstopdf}

\DeclareMathAlphabet\mathbfcal{OMS}{cmsy}{b}{n}

\begin{document}

\title{Long-range correlation-induced effects at high-order harmonic generation on graphene quantum dots}
\author{H.K. Avetissian}
\author{A.G. Ghazaryan}
\author{Kh.V. Sedrakian}
\author{G.F. Mkrtchian}
\thanks{mkrtchian@ysu.am}

\affiliation{Centre of Strong Fields Physics at Research Institute of Physics, Yerevan State University,
Yerevan 0025, Armenia}

\begin{abstract}
This paper focuses on investigating high-order harmonic generation (HHG) in
graphene quantum dots (GQDs) under intense near-infrared laser fields. To
model the GQD and its interaction with the laser field, we utilize a
mean-field approach. Our analysis of the HHG power spectrum reveals fine
structures and a noticeable enhancement in cutoff harmonics due to the
long-range correlations. We also demonstrate the essential role of Coulomb
interaction in determining of harmonics intensities and cutoff position.
Unlike atomic HHG, where the cutoff energy is proportional to the pump wave
intensity, in GQDs the cutoff energy scales with the square root of the
field strength amplitude. A detailed time-frequency analysis of the entire
range of HHG spectrum is presented using a wavelet transform. The analysis
reveals intricate details of the spectral and temporal fine structures of
HHG, offering insights into the various HHG mechanisms in GQDs.
\end{abstract}

\maketitle

\section{Introduction}

After implementation of lasers, there has been a growing interest in
designing and constructing novel materials with exceptional nonlinear
optical properties, due to their potential applications in optoelectronics 
\cite{rosencher2002optoelectronics,grundmann2002nano} and nanophotonics \cite%
{prasad2004nanophotonics,bonaccorso2010graphene}. Along with the second and
third perturbative harmonics \cite{boyd2020nonlinear}, the extreme nonlinear
response such as high-order harmonics generation \cite{corkum1993plasma} and
wave mixing via nonlinear channels are already actual, which is of great
importance in solving various contemporary problems. These include
spectroscopy of attosecond resolution \cite{krausz2009attosecond},
generation of short wavelength coherent radiation \cite%
{avetissian2015relativistic}, recovery of electronic \cite{vampa2015all} and
topological properties of materials \cite%
{luu2018measurement,avetissian2020high,avetissian2022high}, observation of
dynamical Bloch oscillations \cite{luu2015extreme}, Peierls \cite%
{bauer2018high} and Mott \cite{silva2018high} transitions. Therefore, the
design and development of new materials with unique nonlinear optical
properties remains a highly active research field. Among the novel
materials, the carbonbased nanomaterials, such as fullerenes \cite%
{kroto1985c60,fowler2007atlas}, carbon nanotubes \cite{iijima1991helical},
graphene \cite{geim2009graphene}, graphyne \cite{peng2014new}, graphdiyne 
\cite{li2014graphdiyne} are widely used to design nonlinear optical
materials. Carbon nanosystems are attractive for extreme nonlinear optics
because of the presence of copious delocalized $\pi $-electrons in them.
Graphene possessing extensive $\pi $-conjugation, shows extraordinary
nonlinear properties \cite{avetissian2022efficient}. Theoretical works
predicted a strong HHG from fullerenes \cite%
{zhang2005optical,zhang2006ellipticity,zhang2020high,avetissian2021high,avetissian2023disorder}%
. One of the possibilities to manipulate the optical properties of
graphene-based materials is the further decreasing the dimensionality of the
system to obtain graphene nanoribbon or zero dimensional GQD \cite%
{gucclu2014graphene}. The phenomenon of HHG in graphene nanoribbons \cite%
{cox2017plasmon,avetissian2020extreme,zhang2022extended} and GQDs are also
reported \cite{JETP2022high,JN2022high,JETPL2022laser,gnawali2022ultrafast}
and they found the change in nonlinear optical properties by varying the
size, shape, and edges of these systems. The electronic properties of GQDs
are close to atomic and molecular systems. However, because there are only a
limited number of levels, the HHG process cannot be adequately explained by
the typical three-step model \cite{lewenstein1994theory}, where the
continuum serves as the energy-acquiring location. Instead, the process
exhibits similarities to HHG in atomic systems with the both permanent
dipole moments \cite{avetissian2008efficient,avetissian2011coherent} and no
dipole moments \cite{sundaram1990high,plaja1992adiabatic,
lappas1993spectrum,kaplan1994superdressed,gauthey1995role,de1996wavelet,gauthey1997high}%
, where resonance and level dressing have a significant impact. In atomic
systems with a limited number of levels, the cutoff frequency of HHG is not
solely determined by the system's intrinsic energy offset. Rather, it is
influenced by transitions between numerous virtual states that are
manipulated by the amplitude and frequency of the wave field, through the
process of level dressing induced by the strong laser field \cite%
{gauthey1997high,avetissian2011coherent}. The successful model for the
description of GQDs is the tight-binding (TB) one with various
parametrizations. In the homogeneous electric field $\mathbf{E}\left(
t\right) $ according to Peierls substitution \cite{peierls1997selected} the
hopping integral $t_{ij}$ acquire a phase $\left( \mathbf{r}_{i}-\mathbf{r}%
_{j}\right) \int \mathbf{E}\left( t\right) dt$ upon electron tunneling from
site $\mathbf{r}_{i}$ to $\mathbf{r}_{j}$. In the mean-field approximation,
the electron-electron interaction (EEI) alters hopping integrals and causes
them to become non-zero between distant nodes, regardless of their
separation. The Hansen-Bessel formula \cite{ito1993encyclopedic} suggests
that a phase exhibiting large amplitude oscillations is equivalent to high
harmonic oscillations of the effective hopping integrals. Therefore, in case
of HHG in the strong fields, the second-next-nearest, third-next-nearest
hopping, and long-range Coulomb interactions may all have a significant
impact. It is worth noting that in graphene, the second-next-nearest hopping
breaks electron-hole symmetry \cite{kretinin2013quantum}, which is important
for a doped system. However, if one considers an undoped system and neglects
thermal occupations, the HHG process is unaffected by the
second-next-nearest hopping.

Theoretical analyses of HHG in GQD have so far focused on a free electron
model with the next nearest hopping integral \cite{gnawali2022ultrafast} and
short-range Coulomb interactions only \cite%
{JETP2022high,JN2022high,JETPL2022laser}. However, the question of influence
of the long-range correlations on the HHG process and sub-cycle electronic
response in these systems remains unclear. In this study, we develop a
microscopic theory of GQD nonlinear interaction with the strong
electromagnetic radiation that takes into account the long-range hopping
integrals and EEI. Specifically, we investigate two GQDs, shown in Figure 1,
with the same point group symmetry $C_{6v}$ but differing in the number of
atoms -- GQD$_{24}$ and GQD$_{54}$, allowing us to study size effects. Using
the dynamical Hartree-Fock (HF) approximation, we uncover the general and
fundamental structure of the HHG spectrum depending on the long-range
parameters. Our investigations not only provide particular results for GQDs
but also have the potential to be generalized to other systems within this
family.

\begin{figure}[tbp]
\includegraphics[width=0.5\textwidth]{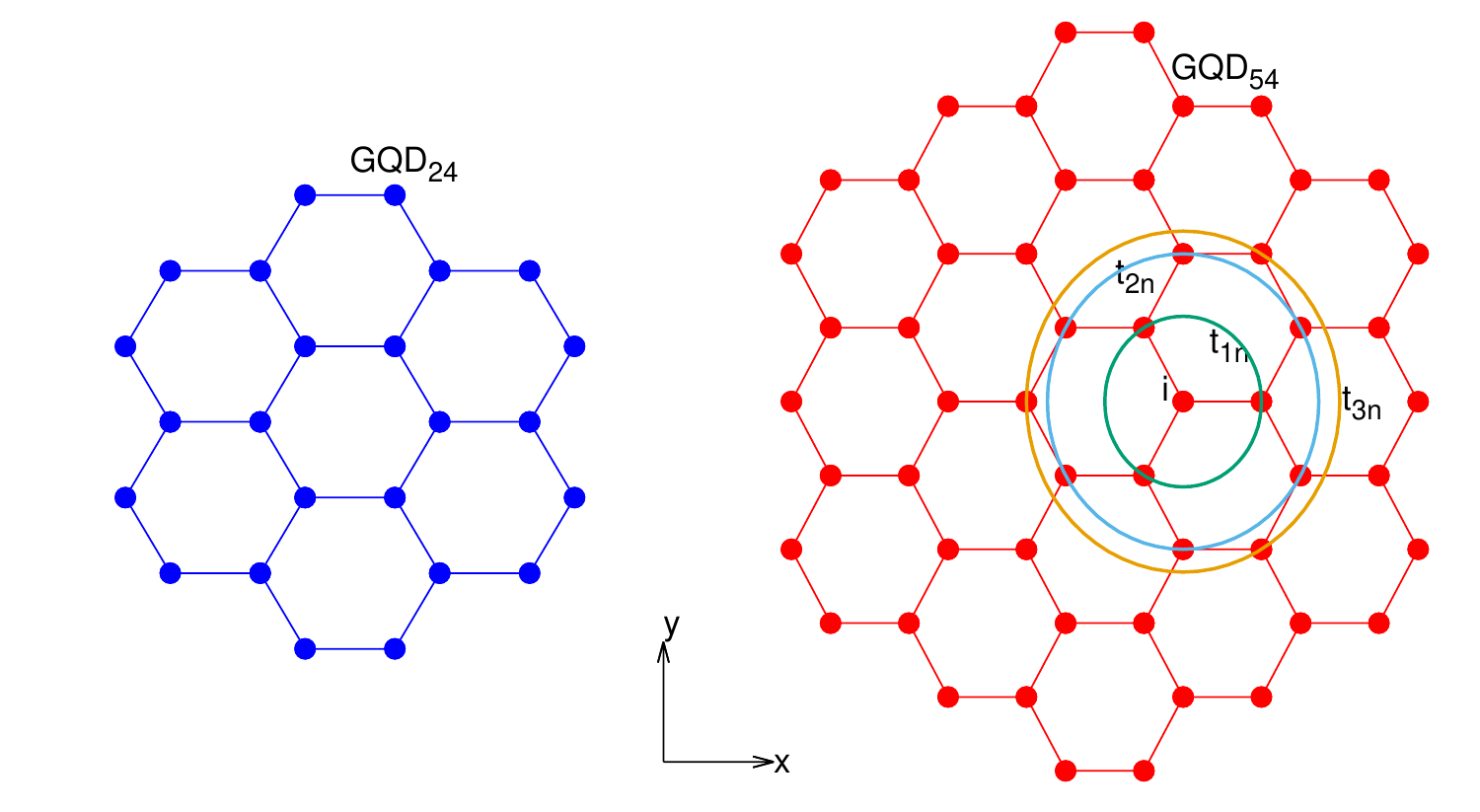}
\caption{Shown here are the schematic structures of GQD$_{24}$ and GQD$_{54}$%
, which share the same point group as graphene. We also display the geometry
used and highlight the nearest-neighbor (3 atoms), next-nearest-neighbor (6
atoms), and third nearest-neighbor (3 atoms) sites, along with their
corresponding hopping integrals.}
\end{figure}

The paper is organized as follows. In Sec. II, the model and the basic
equations are formulated. In Sec. III, we present the main results. Finally,
conclusions are given in Sec. IV.

\section{The model and theoretical methods}

We start by describing the model and theoretical approach. GQD is assumed to
interact with a mid-infrared or visible light laser pulse that excites
electron coherent dynamics. In particular, we consider the two versions of
GQD. Both GQDs which diployed in Fig. 1 are invariant under the inversion
with respect to the center of mass. We assume a neutral GQD, which will be
described in the scope of the TB theory. Hence, the total Hamiltonian reads:%
\textrm{\ }

\begin{equation}
\widehat{H}=\widehat{H}_{\mathrm{TB}}+\widehat{H}_{\mathrm{C}}+\widehat{H}_{%
\mathrm{int}},  \label{1H}
\end{equation}%
where%
\begin{equation}
\widehat{H}_{\mathrm{TB}}=-\sum_{i,j\sigma }t_{ij}c_{i\sigma }^{\dagger
}c_{j\sigma }  \label{2H}
\end{equation}%
is the free GQD TB\ Hamiltonian. Here $c_{i\sigma }^{\dagger }$\ ($%
c_{i\sigma }$) creates (annihilates) an electron with the spin polarization $%
\sigma =\left\{ \uparrow ,\downarrow \right\} $\ at the site $i$\ ($%
\overline{\sigma }$\ is the opposite to $\sigma $\ spin polarization). The
hopping integral, $t_{ij}$, is taken up to third-nearest-neighbor. The
overlap integrals and longer range interactions could also be included but
they are expected to have minor effect on the process considered. For
hopping integral we assume $t_{1n}=2.78\ \mathrm{eV}$ for nearest-neighbor, $%
t_{2n}=0.12\ \mathrm{eV}$ for next-nearest-neighbor, and $t_{3n}=0.068\ 
\mathrm{eV}$ for third nearest-neighbor hopping \cite{kundu2011tight}. The
edges of the GQD are considered to be hydrogen passivated, which has little
effect on the bonds.

The second term in the total Hamiltonian (\ref{1H}) describes the EEI.
Within the HF approximation, the Hamiltonian $\widehat{H}_{\mathrm{C}}$ is
approximated by,%
\begin{equation*}
\widehat{H}_{C}^{HF}=U\sum_{i}\left( \overline{n}_{i\uparrow }-\overline{n}%
_{0i\uparrow }\right) n_{i\downarrow }
\end{equation*}%
\begin{equation*}
+U\sum_{i}\left( \overline{n}_{i\downarrow }-\overline{n}_{0i\downarrow
}\right) n_{i\uparrow }+\sum_{i,j}V_{ij}\left( \overline{n}_{j}-\overline{n}%
_{0j}\right) n_{i}
\end{equation*}%
\begin{equation}
-\sum_{i,j\sigma }V_{ij}c_{i\sigma }^{\dagger }c_{j\sigma }\left(
\left\langle c_{i\sigma }^{\dagger }c_{j\sigma }\right\rangle -\left\langle
c_{i\sigma }^{\dagger }c_{j\sigma }\right\rangle _{0}\right) ,  \label{2s}
\end{equation}%
with on-site and inter-site Coulomb repulsion energies $U$\ and $V_{ij}$,
respectively. The density operator is: $n_{i\sigma }=c_{i\sigma }^{\dagger
}c_{i\sigma }$, and the total electron density for the site $i$\ is: $%
n_{i}=n_{i\uparrow }+n_{i\downarrow }$. Here $\overline{n}_{i\sigma
}=\left\langle c_{i\sigma }^{\dagger }c_{i\sigma }\right\rangle $ and $\rho
_{ji}^{\left( \sigma \right) }=\left\langle c_{i\sigma }^{\dagger
}c_{j\sigma }\right\rangle $. The Coulomb interaction matrix elements can be
obtained from numerical calculations by using Slater $\pi _{z}$ orbitals 
\cite{potasz2010spin,gucclu2014graphene}. Introducing an effective
dielectric constant $\epsilon $ which accounts for the substrate-induced
screening in the 2D nanostructure, we take onsite interaction parameter as $%
U=16.5/\epsilon $ $\ \mathrm{eV}$, $V_{ij}=8.6/\epsilon $ $\mathrm{eV}$ for
nearest-neighbor, $V_{ij}=5.3/\epsilon $ $\mathrm{eV}$ for
next-nearest-neighbor. The longer range Coulomb interaction is taken to be $%
V_{ij}=14.4/\left( \epsilon d_{ij}\right) $ $\mathrm{eV}$, where $d_{ij}$ is
the distance in angstrom between the distant neighbors. In Coulomb
Hamiltonian Eq. (\ref{2s}) the exchange and scattering terms are neglected.
The effective dielectric constant is taken to be $\epsilon =6$ \cite%
{ando2006screening}. Since in the TB\ Hamiltonian we assumed bulk graphene
parameters in the Hartree-Fock Hamiltonian (\ref{2s}) we subtract the
graphene bulk density matrix $\rho _{0ji}^{\left( \sigma \right) }\equiv
\left\langle c_{i\sigma }^{\dagger }c_{j\sigma }\right\rangle _{0}$ ($%
\overline{n}_{0i\sigma }=\rho _{0ii}^{\left( \sigma \right) }$) already
present in the TB term \cite{gucclu2009magnetism,gucclu2014graphene}.

The last term in the total Hamiltonian (\ref{1H}) is the light-matter
interaction part that is described in the length-gauge via the pure scalar
potential,%
\begin{equation}
\widehat{H}_{\mathrm{int}}=e\sum_{i\sigma }\mathbf{r}_{i}\cdot \mathbf{E}%
\left( t\right) c_{i\sigma }^{\dagger }c_{i\sigma },  \label{intH}
\end{equation}%
with the elementary charge $e$, position vector $\mathbf{r}_{i}$, and the
electric field strength $\mathbf{E}\left( t\right) =f\left( t\right) E_{0}%
\hat{\mathbf{e}}\cos \omega t$, with the frequency $\omega $, polarization $%
\hat{\mathbf{e}}$ unit vector, and amplitude $E_{0}$. The wave envelope is
described by the Gaussian function $f\left( t\right) =\exp \left[ -2\ln
2\left( t-t_{m}\right) ^{2}/\mathcal{T}^{2}\right] $, where $\mathcal{T}$
characterizes the pulse duration full width at half maximum, $t_{m}$ defines
the position of the pulse maximum. Note that for the Gaussian envelope the
number of oscillations $N_{s}$ of the field is approximated as $\mathcal{T}%
/T\simeq 0.307N_{s}$, where $T=2\pi /\omega $ is the wave period.

From the Heisenberg equation we obtain evolutionary equations for the
single-particle density matrix $\rho _{ij}^{\left( \sigma \right)
}=\left\langle c_{j\sigma }^{\dagger }c_{i\sigma }\right\rangle $: 
\begin{equation*}
i\hbar \frac{\partial \rho _{ij}^{\left( \sigma \right) }}{\partial t}%
=\sum_{k}\left( \tau _{kj\sigma }\rho _{ik}^{\left( \sigma \right) }-\tau
_{ik\sigma }\rho _{kj}^{\left( \sigma \right) }\right) +\left( V_{i\sigma
}-V_{j\sigma }\right) \rho _{ij}^{\left( \sigma \right) }
\end{equation*}%
\begin{equation}
+e\mathbf{E}\left( t\right) \left( \mathbf{r}_{i}-\mathbf{r}_{j}\right) \rho
_{ij}^{\left( \sigma \right) }-i\hbar \gamma \left( \rho _{ij}^{\left(
\sigma \right) }-\rho _{eij}^{\left( \sigma \right) }\right) ,  \label{evEqs}
\end{equation}%
where 
\begin{equation}
V_{i\sigma }=\sum_{j\alpha }V_{ij}\left( \rho _{jj}^{\left( \alpha \right)
}-\rho _{0jj}^{\left( \alpha \right) }\right) +U\left( \rho _{ii}^{\left( 
\overline{\sigma }\right) }-\rho _{0ii}^{\left( \overline{\sigma }\right)
}\right) ,  \label{Vij}
\end{equation}%
\ and 
\begin{equation}
\tau _{ij\sigma }=t_{ij}+V_{ij}\left( \rho _{ji}^{\left( \sigma \right)
}-\rho _{0ji}^{\left( \sigma \right) }\right)   \label{tauij}
\end{equation}%
are defined via$\ $the$\ $density matrix $\rho _{ij}^{\left( \sigma \right) }
$\ and Coulomb terms. As we see due to the mean field modification hopping
integrals become non-zero between the remote nodes, irrespective of the
distance.

In the Hartree-Fock or mean-field approximation, the higher order
correlation terms are neglected. Calculating the dynamics of the correlation
terms (in the second Born approximation) allows the investigation of
scattering processes which have been introduced in Eq. (\ref{evEqs})
phenomenologically via damping term, assuming that the system relaxes at a
rate $\gamma $\ to the equilibrium $\rho _{eij}^{\left( \sigma \right) }$\
distribution. Optically excited electrons undergo relaxation processes
towards equilibrium through various scattering mechanisms, which include
interactions such as electron-phonon, electron-electron, or
electron-impurity scattering. In the case of graphene-like nanostructures,
both experimental investigations \cite%
{lui2010ultrafast,breusing2011ultrafast} and theoretical analyses \cite%
{hwang2007inelastic,tse2008ballistic} indicate that the predominant
influence on relaxation dynamics arises from electron-electron scatterings
characterized by timescales typically on the order of tens of femtoseconds.
In the context of GQDs, the relaxation rates can be even more pronounced due
to the reduction in dynamic screening effects. Therefore, in the present
study, we have established a relaxation rate of\ $\hbar \gamma =0.2\ \mathrm{%
eV}$.

\section{Results}

After the outlining theoretical setup, we begin by examining the HF
quasi-particle spectrum, which is characterized by a fully occupied valence
band and a completely empty conduction band. The eigenstates and
eigenenergies of the systems under consideration prior to interaction with a
strong laser pulse can be determined self-consistently by iterating Eq. (\ref%
{2s}). To calculate the bulk density matrix, we use a sufficiently large GQD
and place reference GQDs in the middle, as shown in Fig. 2. This approach
allows us to obtain the necessary bulk density matrix elements for the
reference GQDs, which are not influenced by boundary effects. With the help
of the obtained eigenstates $\psi _{\mu }\left( i\right) $ we also calculate
the matrix elements of the transition dipole moment: 
\begin{equation}
\mathbf{d}_{\mu ^{\prime }\mu }=e\sum_{i}\psi _{\mu ^{\prime }}^{\ast
}\left( i\right) \mathbf{r}_{i}\psi _{\mu }\left( i\right) .  \label{dmm'}
\end{equation}%
Beginning the iteration with TB orbitals, the converged results are
presented in Figs. 3 and 4. Intraband and interband transitions in GQD$_{24}$%
\ and GQD$_{54}$\ are analyzed in terms of the energy-difference dependence
of the absolute values of transition dipole moment's matrix elements in Fig.
3. As is seen from Fig. 3, the Coulomb interaction shifts transitions peaks
to higher energies, that is oscillator strengths at higher energies have
relatively larger weight than in the case of free electrons. This effect is
due to the fact that the long range Coulomb interactions give rise to large
hopping integrals between the remote nodes (\ref{tauij}) in the HF
approximation. Eigenenergies with and without EEI for GQD$_{24}$\ and GQD$%
_{54}$\ are presented in Figs. (4a) and (4b), respectively, demonstrating
the effect of EEI on the system's energy spectrum. Compared with the GQD$%
_{24}$, the GQD$_{54}$\ has more nearly degenerated states, and as a
consequence, more channels for the interband and intraband transitions.

\begin{figure}[tbp]
\includegraphics[width=0.3\textwidth]{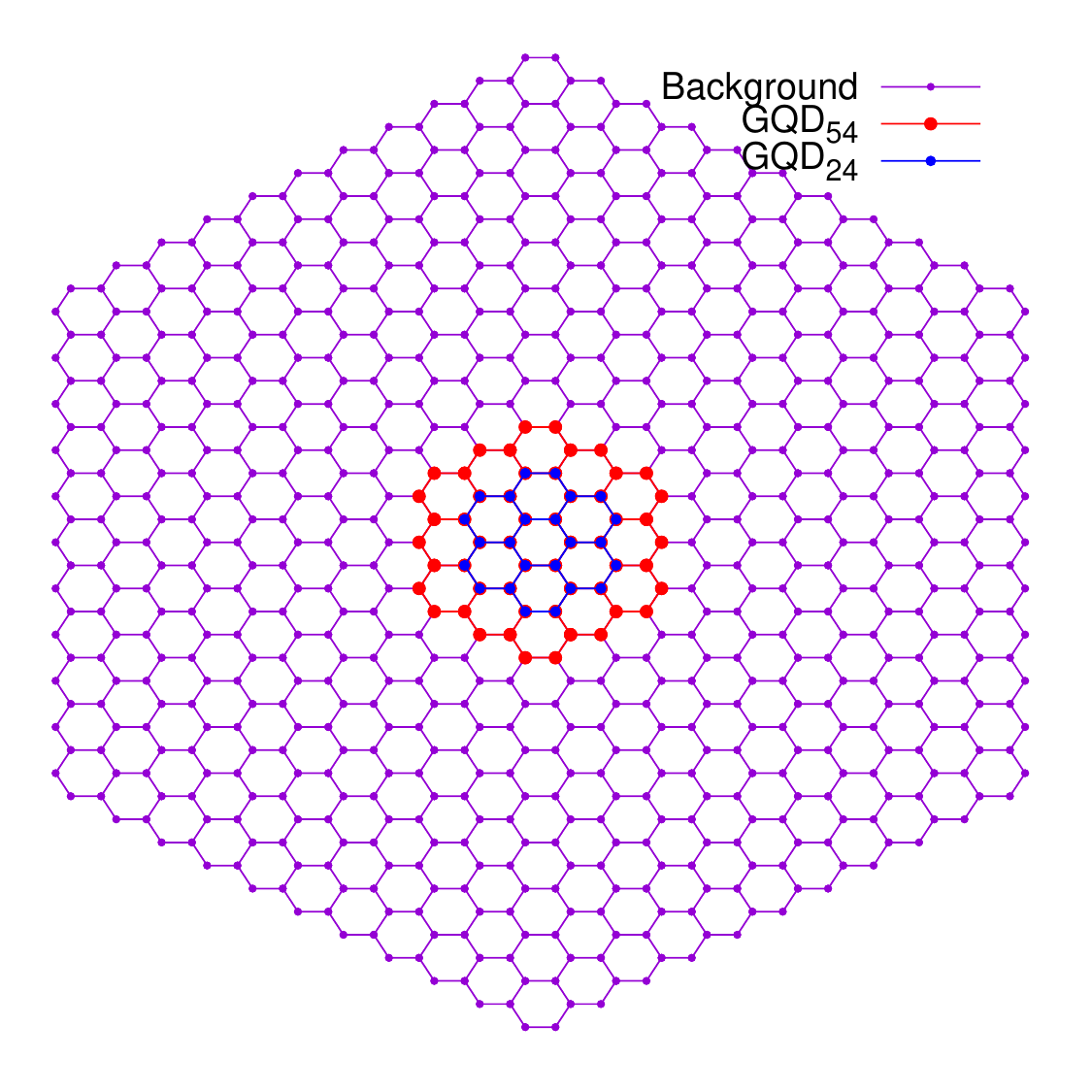}
\caption{To determine the bulk density matrix, we employ GQD$_{726}$ and
position the reference GQDs at its center. The density matrix elements of GQD%
$_{726}$ are utilized as a bulk in the Hartree-Fock Hamiltonian (\protect\ref%
{2s}).}
\end{figure}

\begin{figure*}[tbp]
\includegraphics[width=1.0\textwidth]{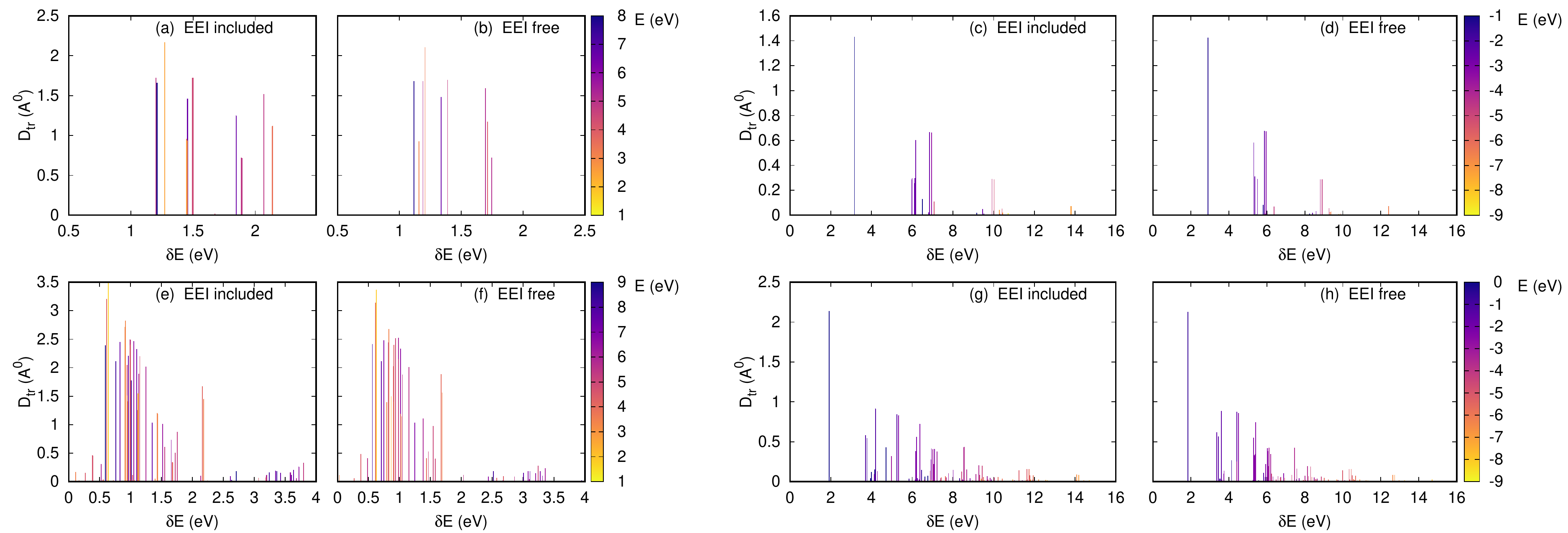}
\caption{Absolute values of transition dipole moment matrix elements are
presented for GQD$_{24}$ (a, b, c, d) and GQD$_{54}$ (e, f, g, h), showing
the energy dependence of intraband (a, b, e, f) and interband transitions
(c, d, g, h), with and without EEI. For intraband trazitions it is shown
only conduction band, since similar picture we have for valence band. The
color boxes show energy ranges of the corresponding bands. }
\label{fig3m}
\end{figure*}

\begin{figure}[tbp]
\includegraphics[width=0.42\textwidth]{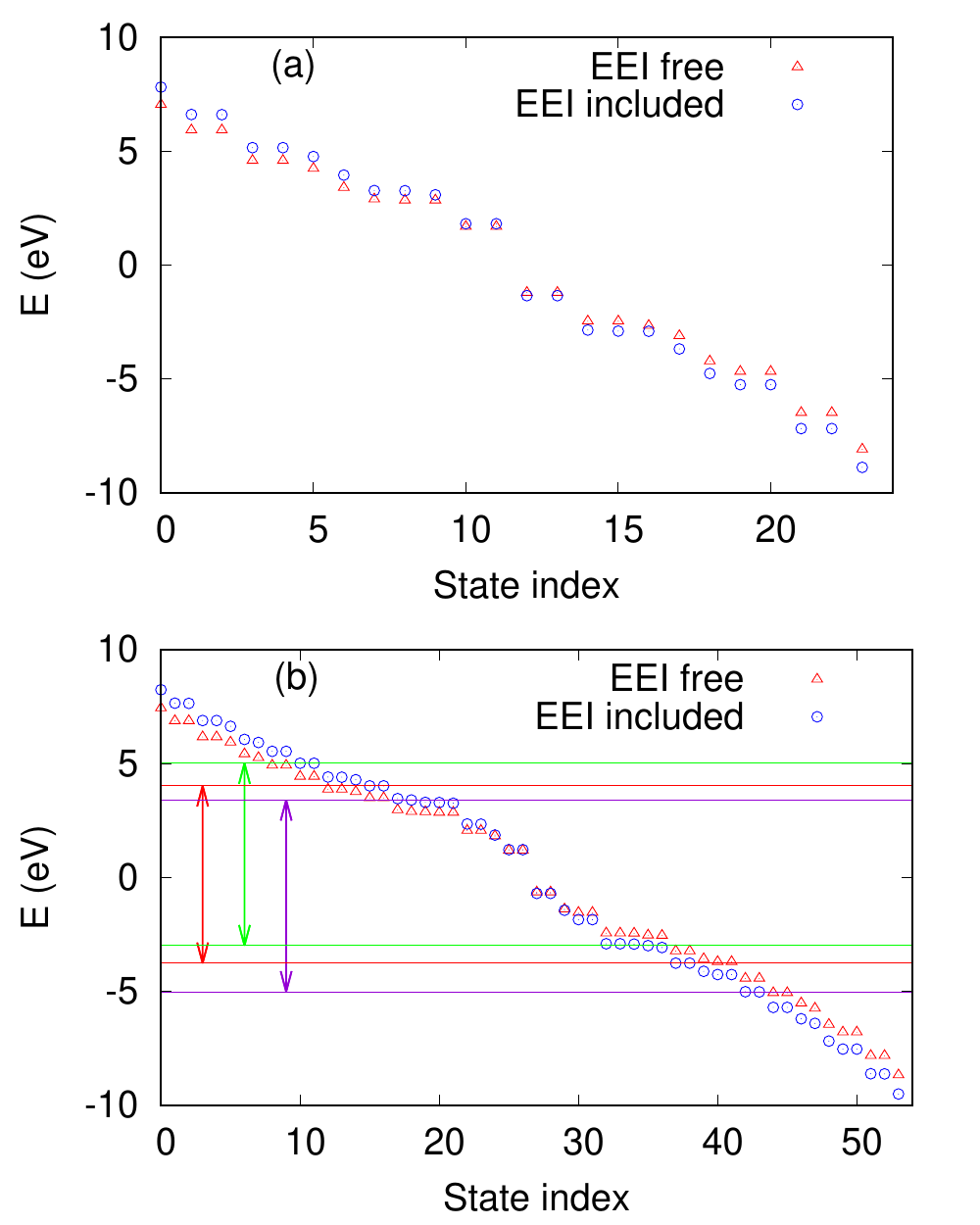}
\caption{Eigenenergies with and without EEI for GQD$_{24}$ and GQD$_{54}$
are presented in (a) and (b), respectively. In (b) it is shown also resonant
levels with corresponding transitions responsible for five photon resonance.}
\end{figure}

The HHG spectrum is obtained by taking the Fourier transform $\mathbf{a}%
\left( \Omega \right) $ of the dipole acceleration $\mathbf{a}\left(
t\right) =d^{2}\mathbf{d}/dt^{2}$, where the dipole moment is defined as 
\begin{equation}
\mathbf{d}\left( t\right) =e\sum_{i\sigma }\mathbf{r}_{i}\rho _{ii}^{\left(
\sigma \right) }\left( t\right) .  \label{dmoment}
\end{equation}

Although we perform our calculations in the coordinate basis, for physical
insight it is useful to consider also the dynamics in the energetic
representation where there are two contributions to the dipole moment: the
transitions of electrons/holes within conduction and valence bands, and the
creation of electron-hole pairs (transitions from occupied molecular
orbitals to unoccupied ones) followed by their recombination. To distinguish
intraband and interband contributions within the dipole acceleration
spectrum, we need to perform a basis transformation using the formula:%
\begin{equation}
\rho _{ij}=\sum_{\mu ^{\prime }}\sum_{\mu }\psi _{\mu ^{\prime }}^{\ast
}\left( j\right) \varrho _{\mu \mu ^{\prime }}\psi _{\mu }\left( i\right) ,
\label{2}
\end{equation}%
where $\psi _{\mu }\left( i\right) $\ represents the Hartree-Fock orbitals,
and $\rho _{\mu \mu ^{\prime }}$\ is the density matrix in the energetic
representation. The inverse transformation is given by: $\varrho _{\mu \mu
^{\prime }}=\sum_{i}\sum_{j}\psi _{\mu }^{\ast }\left( i\right) \rho
_{ij}\psi _{\mu ^{\prime }}\left( j\right) $. Since our
consideration does not involve spin effects, we omit the spin index and
multiply our results by the degeneracy factor of $2$. In this representation 
$\varrho _{\mu \mu }$\ is the population of state with energy $\varepsilon
_{\mu }$, while nondiagonal elements of the density matrix $\varrho _{\mu
\mu ^{\prime }}$\ characterize coupling of levels (i.e., the appearance of
coherence). Using Eqs. (\ref{dmm'}) and (\ref{2}), we can express the dipole
moment (\ref{dmoment}) as:%
\begin{equation}
\mathbf{d}\left( t\right) =\mathbf{d}_{\mathrm{intra}}\left( t\right) +%
\mathbf{d}_{\mathrm{inter}}\left( t\right) ,  \label{4}
\end{equation}%
where%
\begin{equation}
\mathbf{d}_{\mathrm{intra}}\left( t\right) =2\sum_{\mu ,\mu ^{\prime
}=N/2}^{N-1}\varrho _{\mu \mu ^{\prime }}\left( t\right) \mathbf{d}_{\mu
^{\prime }\mu }+2\sum_{\mu ,\mu =0}^{N/2-1}\varrho _{\mu \mu ^{\prime
}}\left( t\right) \mathbf{d}_{\mu ^{\prime }\mu },  \label{dintra}
\end{equation}%
represents the intraband part of the dipole moment, and%
\begin{equation}
\mathbf{d}_{\mathrm{inter}}\left( t\right) =2\sum_{\mu ^{\prime
}=N/2}^{N-1}\sum_{\mu =0}^{N/2-1}\varrho _{\mu \mu ^{\prime }}\left(
t\right) \mathbf{d}_{\mu ^{\prime }\mu }+\mathrm{c.c.},  \label{dinter}
\end{equation}%
represents the interband part. We preferred for calculations in the
coordinate basis due to computational efficiency that is approximately one
order of magnitude faster than in the energetic representation.

To compute the harmonic signal, we use the Fourier transform 
\begin{equation*}
\mathbf{a}\left( \Omega \right) =\int_{0}^{\mathcal{T}}\mathbf{a}\left(
t\right) e^{i\Omega t}W\left( t\right) dt,
\end{equation*}%
\begin{figure}[tbp]
\includegraphics[width=0.5\textwidth]{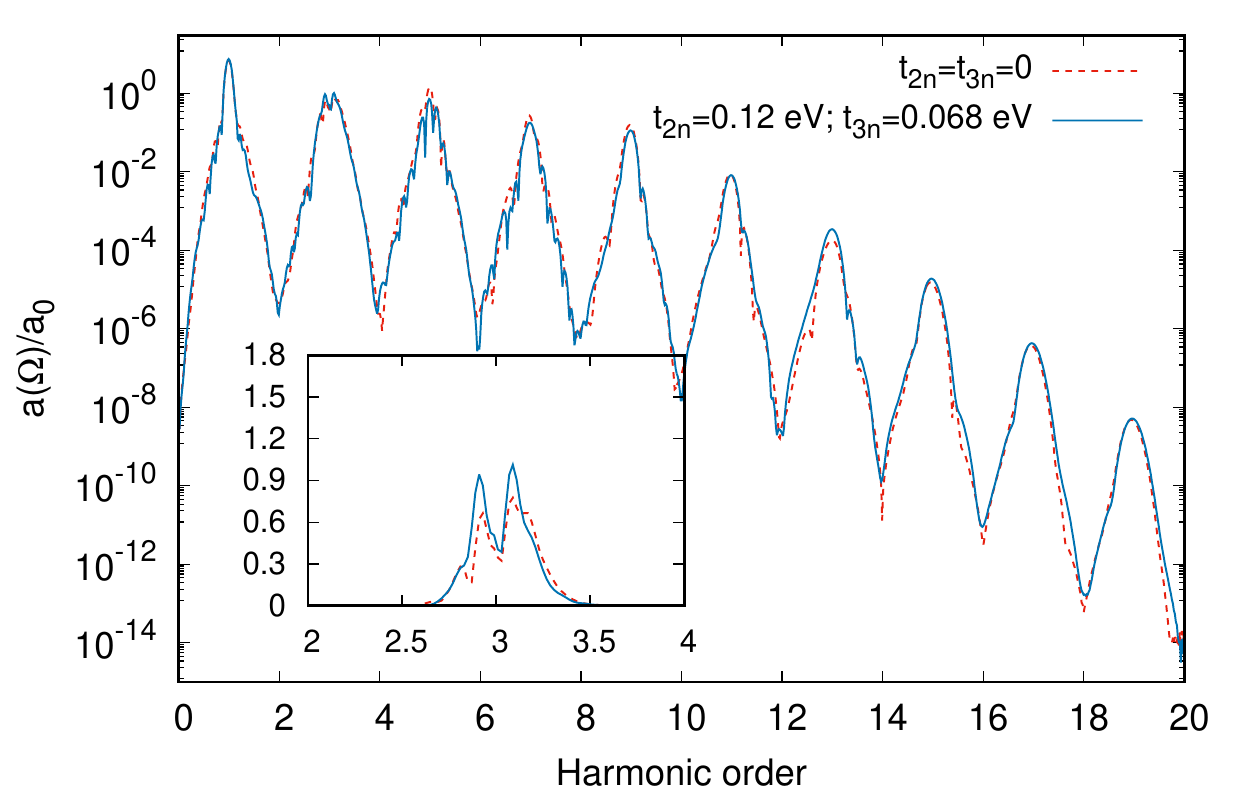}
\caption{The HHG spectra for GQD$_{54}$ in logarithmic scale via the
normalized dipole acceleration Fourier transformation $a\left( \Omega
\right) /a_{0}$ (in arbitrary units) with and without long-range hopping
integrals. The wave amplitude is taken to be $E_{0}=0.5\ \mathrm{V/\mathring{%
A}}$. The inset shows HHG spectra in the linear scale near the 3rd harmonic.
The EEI is switched off.}
\end{figure}
where $W\left( t\right) $ is a window function that suppresses small
fluctuations and reduces the overall background noise of the harmonic signal 
\cite{zhang2018generating}. We choose the pulse envelope $f\left( t\right) $
as the window function. The excitation is performed using a Ti:sapphire
laser with a wavelength of $780\ \mathrm{nm}$, an excitation frequency of $%
\omega =1.59\ \mathrm{eV}/\hbar $, which is comparable to the gaps of GQD$%
_{24}$ ($3.16\ \mathrm{eV}$) and GQD$_{54}$ ($1.92\ \mathrm{eV)}$. For all
further calculations we assume that the wave is linearly polarized with a
polarization unit vector $\hat{\mathbf{e}}=\left\{ 1,0\right\} $, and the
pulse duration $\mathcal{T}$ is set to $20\ \mathrm{fs}$, corresponding to
approximately $25$ oscillations ($N_{s}\simeq 25$). To ensure a smooth
turn-on of the interaction, we position the pulse center at $t_{m}=25T$. For
convenience, we normalize the dipole acceleration by the factor $a_{0}=e%
\overline{\omega }^{2}\overline{d},$ where $\overline{\omega }=1\ \mathrm{eV}%
/\hbar $ and $\overline{d}=1\ \mathrm{\mathring{A}}$. The power radiated at
a given frequency is proportional to $\mathbf{a}^{2}\left( \Omega \right) $.
We perform the time integration of Eq. (\ref{evEqs}) using the eighth-order
Runge-Kutta algorithm.

To begin with, we examine the effect of the next-nearest-neighbor and
third-nearest-neighbor hopping integrals on the HHG spectra, with the EEI
turned off for simplicity. Figure 5 illustrates a comparison of the
relative\ HHG spectra for GQD$_{54}$ with and without long-range hopping
integrals. It is evident from this figure that the long-range hopping
integrals have a measurable impact on the HHG spectra. This influence is
more pronounced in the linear scale, as shown in the inset for the 3rd
harmonic, where we observe a difference of up to 50\%. 
\begin{figure}[tbp]
\includegraphics[width=0.5\textwidth]{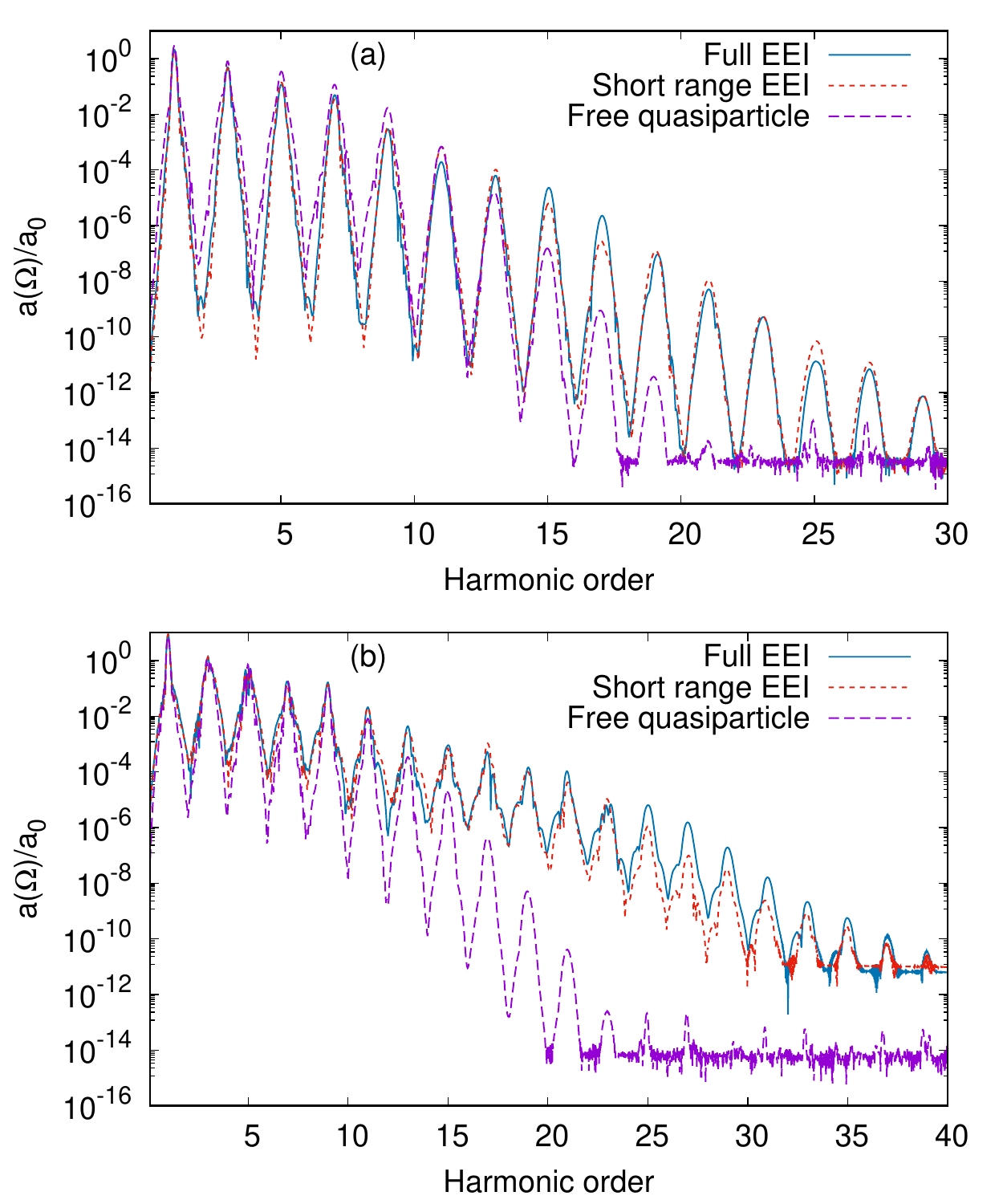}
\caption{The relative HHG spectra for GQD$_{24}$ (a) and for GQD$_{54}$ (b)
via the normalized dipole acceleration Fourier transformation $a\left(
\Omega \right) /a_{0}$ with and without EEI. The wave amplitude is taken to
be $E_{0}=0.5\ \mathrm{V/\mathring{A}}$. }
\end{figure}

Then, we will examine the impact of Coulomb interaction on the HHG spectra.
The HHG spectra are compared for three different scenarios in Fig. 6: when
the full Coulomb effects are considered, when the long-range Coulomb
interaction is eliminated, and when there are no Coulomb effects present,
and the quasiparticles are free. The inclusion of the Coulomb interaction
leads to several noteworthy characteristics in the HHG spectra: (a) the most
prominent feature is a substantial increase in the HHG signal by several
orders of magnitude in the midplateau and near the cutoff regime compared to
the case of free quasiparticles. (b) The cutoff frequency is significantly
enhanced. (c) In the vicinity of the cutoff regime, the spectra are
distinguished by featureless peaks, while for low harmonics there is a
multiple-peak splitting pattern in each main harmonic peak. These fine
structures are entirely reproducible and convey crucial information about
the electron quantum dynamics and the underlying mechanism which will be
discussed later. To observe the effects of the EEI on HHG process in an
actual experiment, it is essential to have the ability to manipulate the
strength of the Coulomb interaction. In most experimental setups, graphene
nanostructures are situated on a substrate, which introduces a screening
effect on the Coulomb interaction, typically reducing it to about $%
1/\epsilon $\ of its original strength, where $\epsilon $\ represents the
substrate's dielectric constant. Specifically, when using a substrate like
SiO$_{2}$, the Coulomb interaction remains moderate. However, by introducing
substrates in contact with liquids of high dielectric constant, like ethanol
($\epsilon \approx 13$), it is possible to significantly enhance the
background dielectric constant. Consequently, in the experiment, one can
vary $\epsilon $, thereby tuning the Coulomb interaction to investigate its
impact on HHG process in GQDs.

The significant enhancement in the HHG signal can be explained by the strong
modification of hopping integrals (\ref{tauij}) and the resulting level
dressing (as shown in Fig. 7) due to the mean field effect. This is
supported by the observation that when the long-range Coulomb interaction is
turned off, the effect is weakened, and these features are more noticeable
in GQD$_{54}$. Another indication is that the maximum harmonic order
generated through transitions between the real energy levels, as clear from
Fig. 4, is only up to the 9th order, while the enhancement occurs in higher
harmonics generated through transitions between the virtual levels. As an
example, Fig. 7 illustrates the population of the state with an energy of $%
2.34$ $\mathrm{eV}$ as a function of time and its Fourier transform. It is
evident from the figure that there are rapid oscillations in the level
populations that include high harmonics of the fundamental frequency. 
\begin{figure}[tbp]
\includegraphics[width=0.5\textwidth]{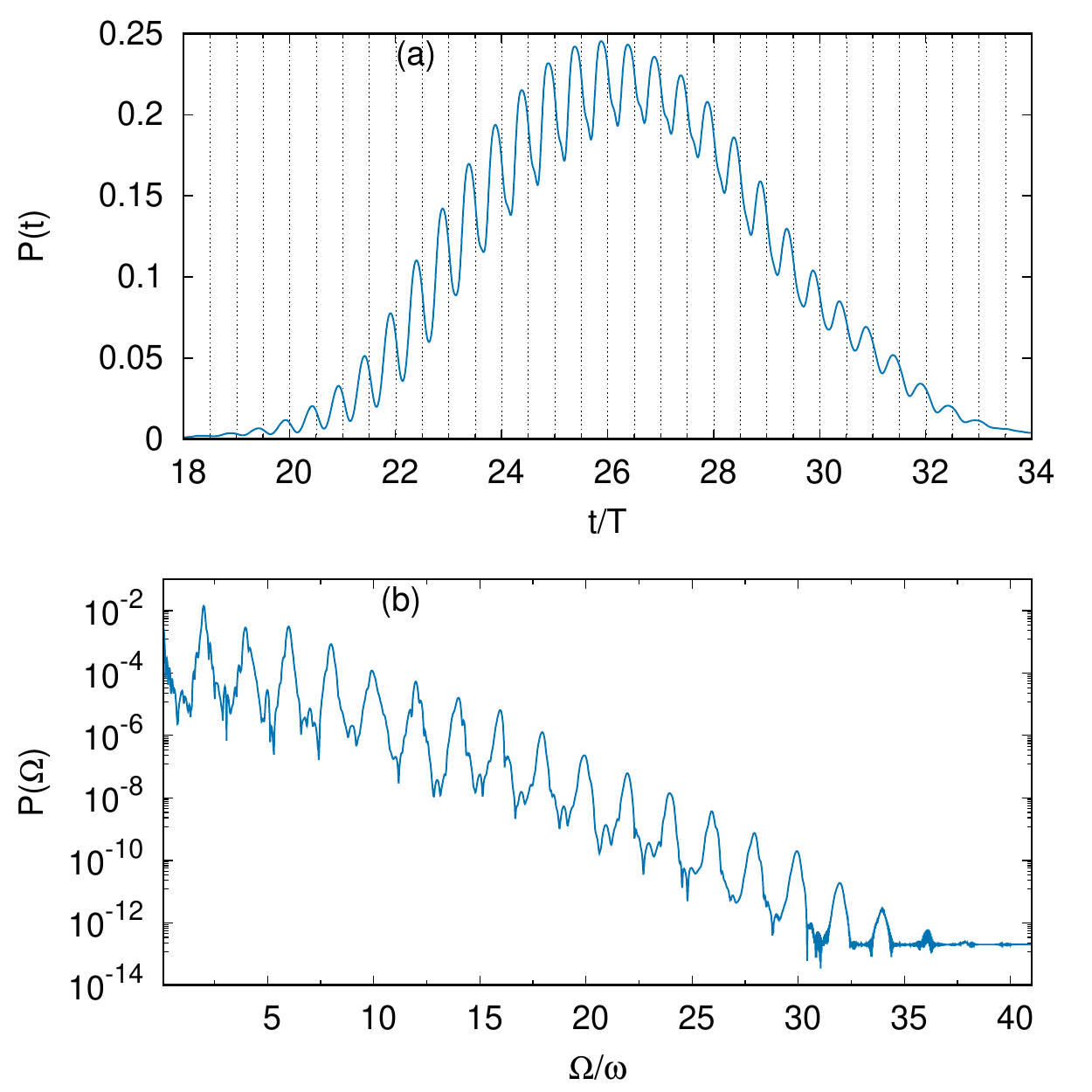}
\caption{(a) Population of the state with an energy of 2.34 $\mathrm{eV}$
versus time (optical cycles) for GQD$_{54}$ and its Fourier transform (b).
Laser parameters same as those in Fig. 5.}
\end{figure}

Another notable aspect of the HHG signals in GQDs is their dependence on the
size of the dot. The HHG signals per particle for GQD$_{24}$ and GQD$_{54}$
are compared in Fig. 8. As demonstrated, there is a significant increase in
the HHG signal for GQD$_{54}$, a result also observed for triangular GQDs
according to previous studies \cite{JETP2022high}. This enhancement may be
attributed to the density of states, which is indirectly reflected in Fig. 3
via the transition dipole moments. This figure reveals that GQD$_{54}$ has
substantially more transition channels than GQD$_{24}$. As the size of the
dot increases, this effect reaches saturation \cite{JETP2022high}. We have
made also calculations for larger dots -- GQD$_{150}$\ and GQD$_{216}$\ and
compared the spectra in the Fig. 8. As is seen, already for GQD$_{150}$\ the
overall enhancement of HHG yield per particle due to the size of the dot is
absent.

\begin{figure}[tbp]
\includegraphics[width=0.5\textwidth]{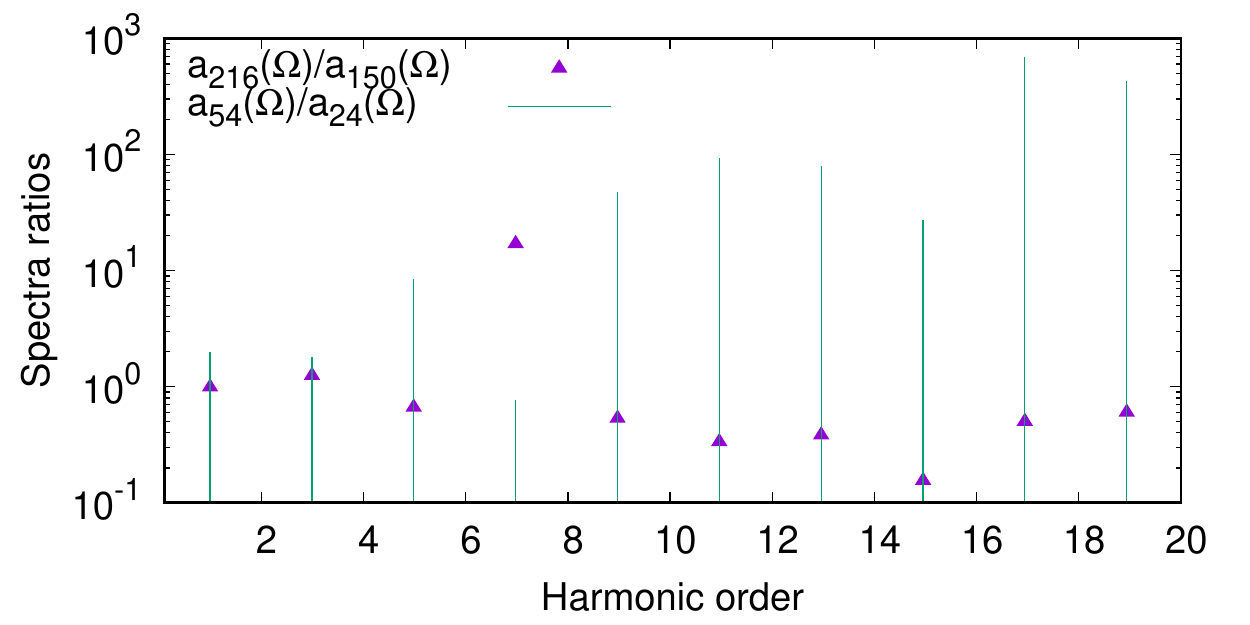}
\caption{The ratio of HHG signals per particle for GQD$_{54}$ to that for GQD%
$_{24}$, using the same laser parameters as in Figure 5. By the triangles it
is shown the ratio of HHG signal per particle for GQD$_{216}$ to that for GQD%
$_{150}$.}
\end{figure}

To investigate the underlying causes of the detailed spectral structures
observed in the HHG spectra Fig. 6, we utilize a wavelet transform \cite%
{tong2000probing} of the dipole acceleration to conduct a time-frequency
analysis. We perform the Morlet transform ($\sigma =20$) of the dipole
acceleration: 
\begin{equation}
\mathbf{a}\left( t,\Omega \right) =\sqrt{\frac{\Omega }{\sigma }}%
\int_{0}^{\tau }dt^{\prime }\mathbf{a}\left( t\right) e^{i\Omega \left(
t^{\prime }-t\right) }e^{-\frac{\Omega ^{2}}{2\sigma ^{2}}\left( t^{\prime
}-t\right) ^{2}}.  \label{wavelet}
\end{equation}

\begin{figure}[tbp]
\includegraphics[width=0.48\textwidth]{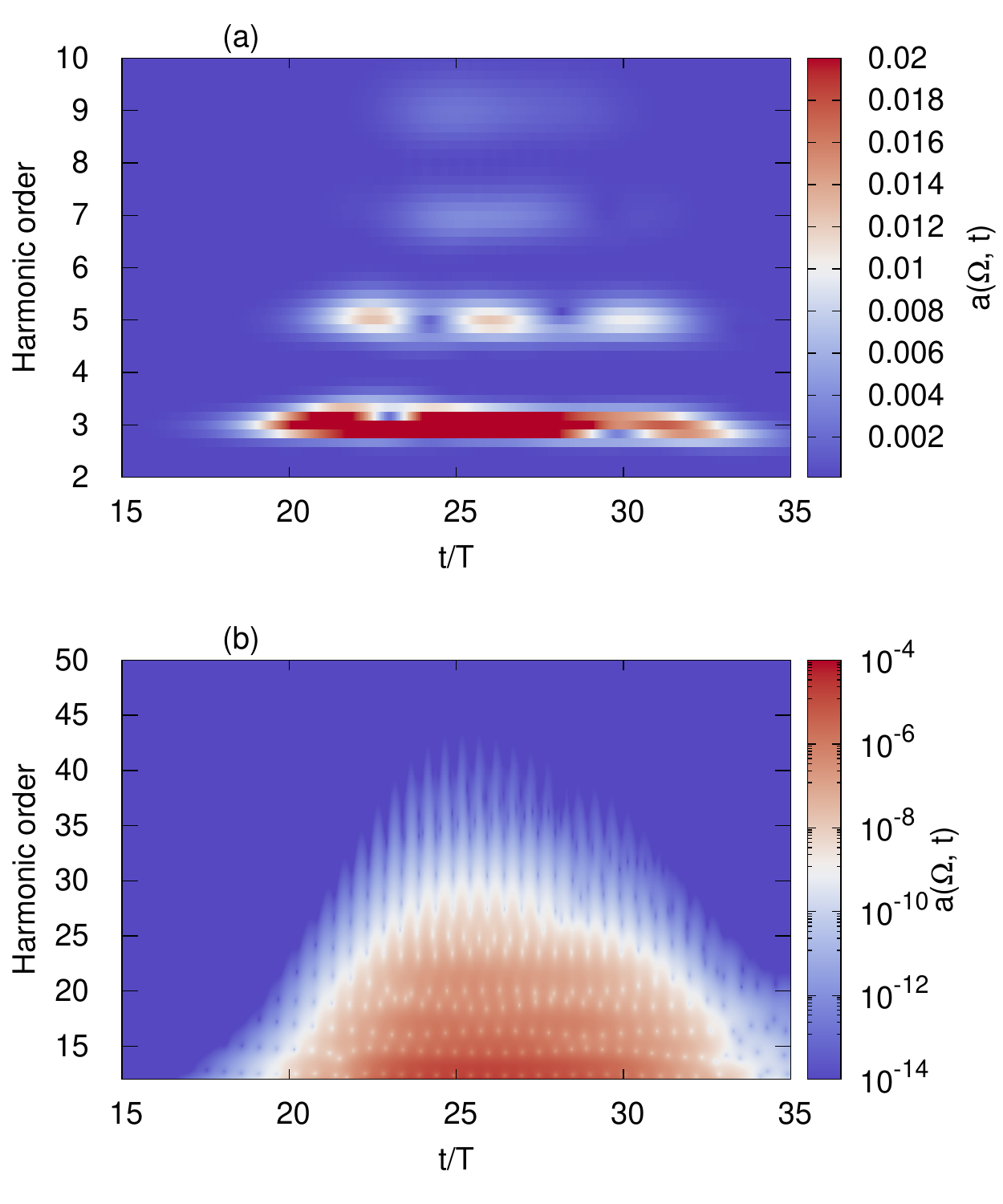}
\caption{The spectrogram (color box in arbitrary units) of the HHG process
via the wavelet transform of the dipole acceleration for GQD$_{54}$: (a) for
the low harmonics and (b) for the high harmonics. It is shown $\left\vert 
\mathbf{a}\left( t,\Omega \right) \right\vert $ in a time interval where the
wave's amplitude is considerable. The laser parameters correspond to Fig. 6.
The similar picture we have for GQD$_{24}$.}
\end{figure}

The wavelet transform of the dipole acceleration provides insight into the
origin of the HHG spectral fine structures. Figure 9 illustrates the
absolute values of the time-frequency profiles of the dipole acceleration,
which have been obtained using the laser parameters presented in Fig. 6 for
GQD$_{54}$. This plot shows remarkable details of the spectral and temporal
structures. Notably, the time profiles for the low harmonics Fig. 9(a)
exhibit a fairly smooth variation with respect to time for a given
frequency. Conversely, for high harmonics near the cutoff region Fig. 9(b),
the most distinct feature is the rapid-burst time profiles. These bursts
occur with a period of $0.5T$, which supports the level dressing model of
HHG, since the time profile resembles the population behavior depicted in
Fig. 7. As an example, the time profile of the harmonic (H23) is represented
in Fig. 10. It demonstrates the occurrence of two bursts during the each
optical cycle.

\begin{figure}[tbp]
\includegraphics[width=0.5\textwidth]{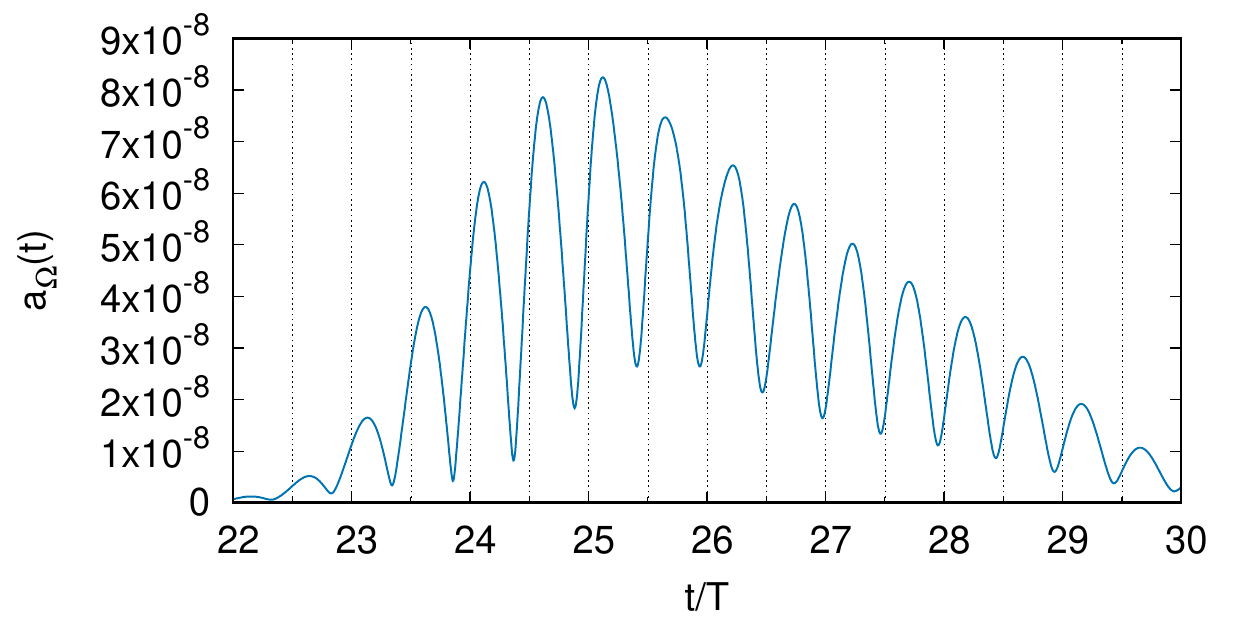}
\caption{The time profile of the 23rd harmonic of GQD$_{54}$. obtained from
cross sections of the time-frequency spectrum in Fig. 9(b). }
\end{figure}

\begin{figure}[tbp]
\includegraphics[width=0.5\textwidth]{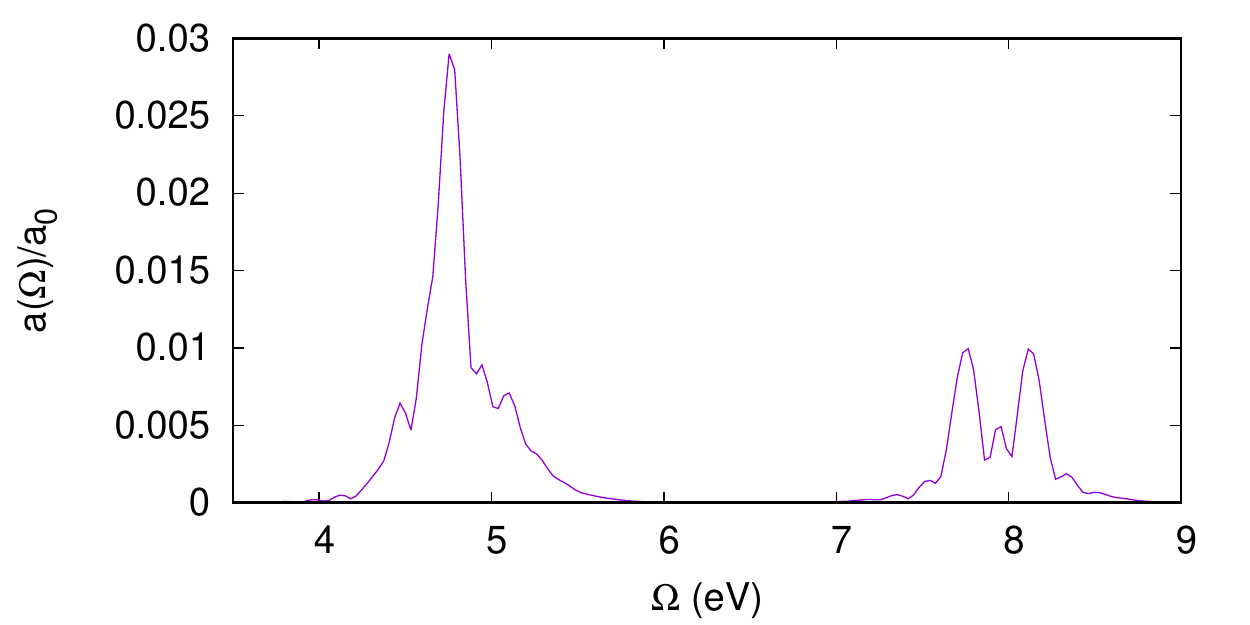}
\caption{The HHG spectrum of GQD$_{54}$ is presented in the linear scale in
the vicinity of the 3rd and 5th harmonics, showcasing the intricate fine
structure of the peaks. The laser parameters used in this analysis
correspond to those presented in Figure 6.}
\end{figure}

Moving on, we will now discuss the multiple-peak splitting patterns observed
for the low harmonics. Figure 11 presents the HHG spectrum of GQD$_{54}$,
highlighting the detailed fine structure of the peaks near the 3rd ($\hbar
\Omega \simeq $ $4.8\ \mathrm{eV}$) and $5$th harmonics ($\hbar \Omega
\simeq $ $8\ \mathrm{eV}$). When we examine Fig. 3(g), we observe that the
dipole moments for interband transitions have peaks around these
frequencies, indicating the possibility of multiphoton resonant transitions.
The sidebands near the $3$rd and $5$th harmonics correspond to these
resonant transitions. This is further supported by the time profiles
obtained from cross sections of the time-frequency spectrum in Fig. 9(a)
around these frequencies. The results for the 3rd and 5th harmonics are
presented in Figures 12 and 13, respectively. In Fig. 12, the profiles for
the $3$rd harmonic and the two sidebands exhibit striking differences,
suggesting distinct mechanisms for their generation. The time profile of the
3rd harmonic closely follows the envelope of the laser pulse, with only
minor modifications attributed to nearby resonances. In contrast, the time
profiles of the sidebands exhibit a relatively flat behavior, indicating
their origin in multiphoton resonant bound-bound transitions. For
nearly resonant transitions, the product $e^{in\omega t}\varrho _{\mu \mu
^{\prime }}\left( t\right) ,$\ which defines time profile of the $n$th
harmonic according to Eq. (\ref{dinter}), has a non-zero average value over
the course of a laser cycle, and it varies slowly within the time frame of
that cycle. Consequently, this leads to the observed flat time profiles of
the sidebands. In the presence of strong laser fields, one should also take
into account the dynamic Stark shifts of energy levels $S_{\mu }\left(
t\right) $, which may become crucial \cite{trallero2005coherent}. These
shifts, which are proportional to the laser intensity profile $%
E_{0}^{2}f^{2}\left( t\right) $, have the capability to adjust otherwise
non-resonant energy levels into resonance conditions, i.e., $\varepsilon
_{\mu }+S_{\mu }\left( t\right) -\varepsilon _{\mu ^{\prime }}-S_{\mu
^{\prime }}\left( t\right) \approx n\omega $\ or vice versa. In the context
of the five-photon resonance transitions illustrated in Fig. 4(b), this
situation arises with three pairs of nearly resonant transitions subjected
to dynamic Stark shift. The time profile of the resulting triple structure
near the $5$th harmonic (as shown in Fig. 13) distinctly displays
oscillations occurring approximately every five optical cycles. This effect
is due to the fact that throughout the interaction of GQDs with the laser
field, populations from valence band states are transferred to excited
conduction band states via dynamic Stark-shifted resonances. These
resonances can emerge multiple times during the pulse \cite%
{jones1995interference}, as the dynamic Stark shift can be either positive
or negative, depending on the specific energy state. Notably, the peaks in
the time profile of the $5$th harmonic correspond to the peaks of levels'
coherence $\varrho _{\mu \mu ^{\prime }}\left( t\right) $. \ It should be
noted that in the midplateau domain of harmonics, we have interplay between
the intraband (\ref{dintra}) and interband (\ref{dinter}) emissions. 
\begin{figure}[tbp]
\includegraphics[width=0.5\textwidth]{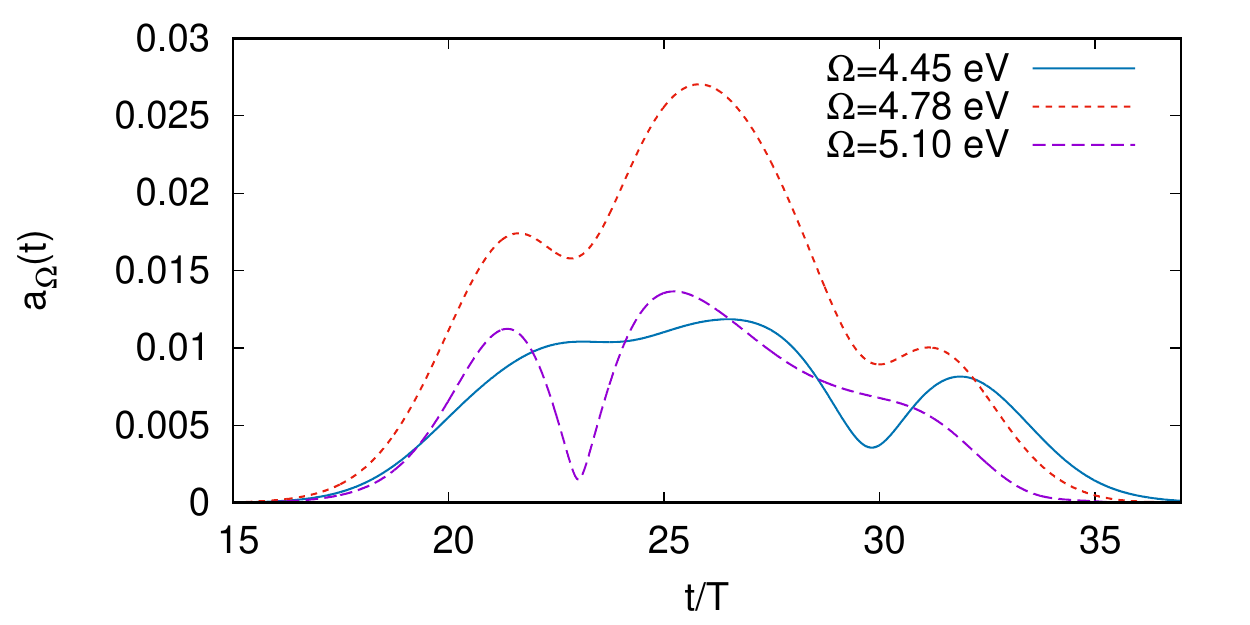}
\caption{The time profiles of the 3rd harmonic and sidebands. Laser
parameters are the same as those in Fig. 9.}
\end{figure}
\begin{figure}[tbp]
\includegraphics[width=0.5\textwidth]{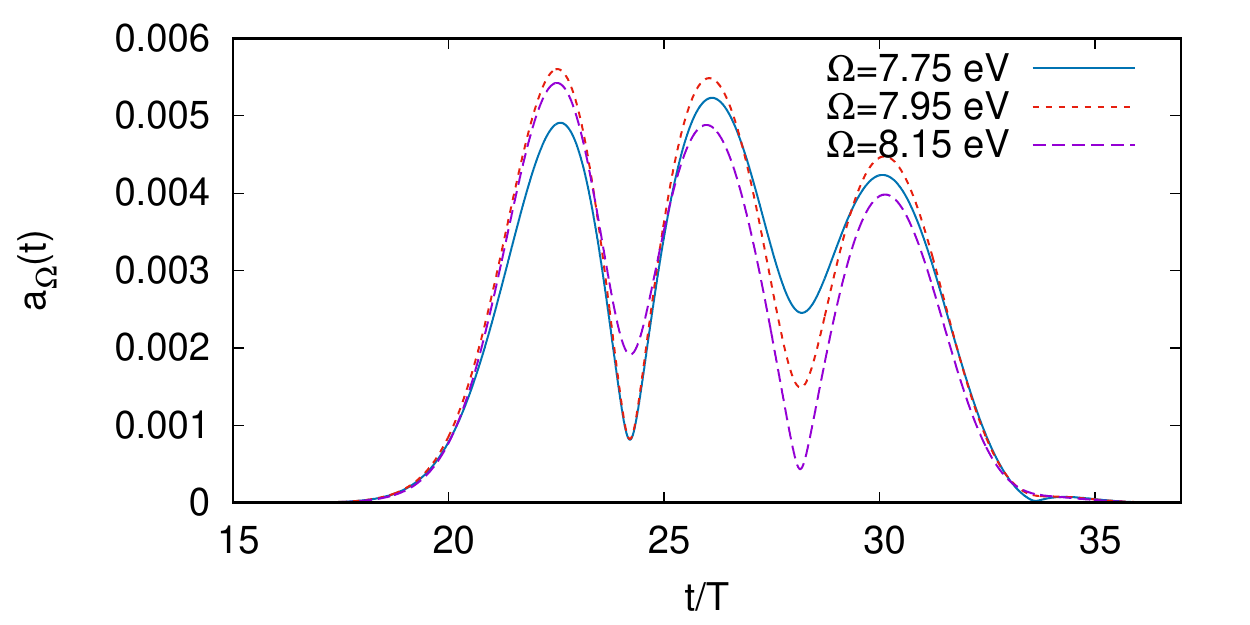}
\caption{The time profiles of the triple structure near the 5th harmonic.
Laser parameters are the same as those in Fig. 9.}
\end{figure}

Next we investigate the dependence of cutoff frequency on the pump wave
intensity by analyzing the HHG spectra for the different intensities. The
dependence of the HHG spectra on the wave field amplitude for both GQDs is
shown in Fig. 14. A significant nonlinear dependence of the mid-plateau and
near-cutoff harmonics on the pump wave amplitude is observed in Fig. 14. In
addition, unlike atomic HHG \cite{lewenstein1994theory}, where the cutoff
energy is proportional to the square of the field strength amplitude, the
cutoff energy in our case scales with the square root of the field strength
amplitude. This trend becomes evident when observing the inset in the lower
panel of Figure 14, wherein we depict the relationship between the cutoff
harmonics and the wave field amplitude ($E_{0}$). The plotted data is
accompanied by a fitting function of the form $\sqrt{E_{0}}$. Within the
range of field strengths spanning up to $0.8\ \mathrm{V/\mathring{A}}$, a
remarkable alignment between the data and the scaling function is observed,
indicative of a strong approximation. Nevertheless, it's noteworthy that a
saturation phenomenon becomes apparent for higher intensity pump waves.

\begin{figure}[tbp]
\includegraphics[width=0.5\textwidth]{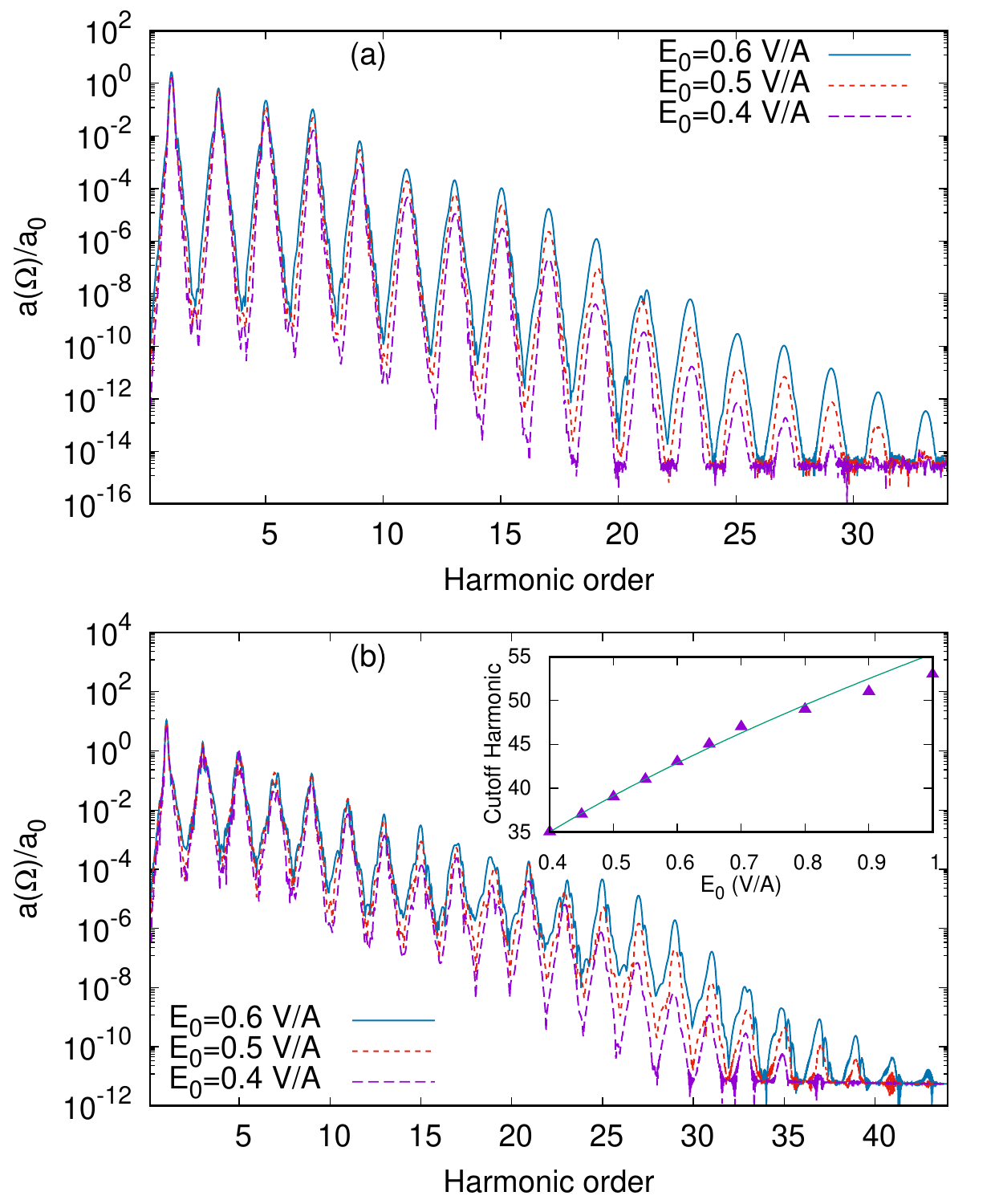}
\caption{The dependencies of the HHG spectra on the wave field amplitude are
illustrated GQD$_{24}$ (a) and for GQD$_{54}$ (b) using the normalized
dipole acceleration Fourier transformation, $a\left( \Omega \right) /a_{0}$,
plotted on a logarithmic scale. The inset in (b) shows the dependence of 
\textrm{cutoff harmonic on the wave field amplitude. The solid line is a
fitting function of the form }$\protect\sqrt{E_{0}}$\textrm{. }}
\label{987}
\end{figure}

\begin{figure}[tbp]
\includegraphics[width=0.5\textwidth]{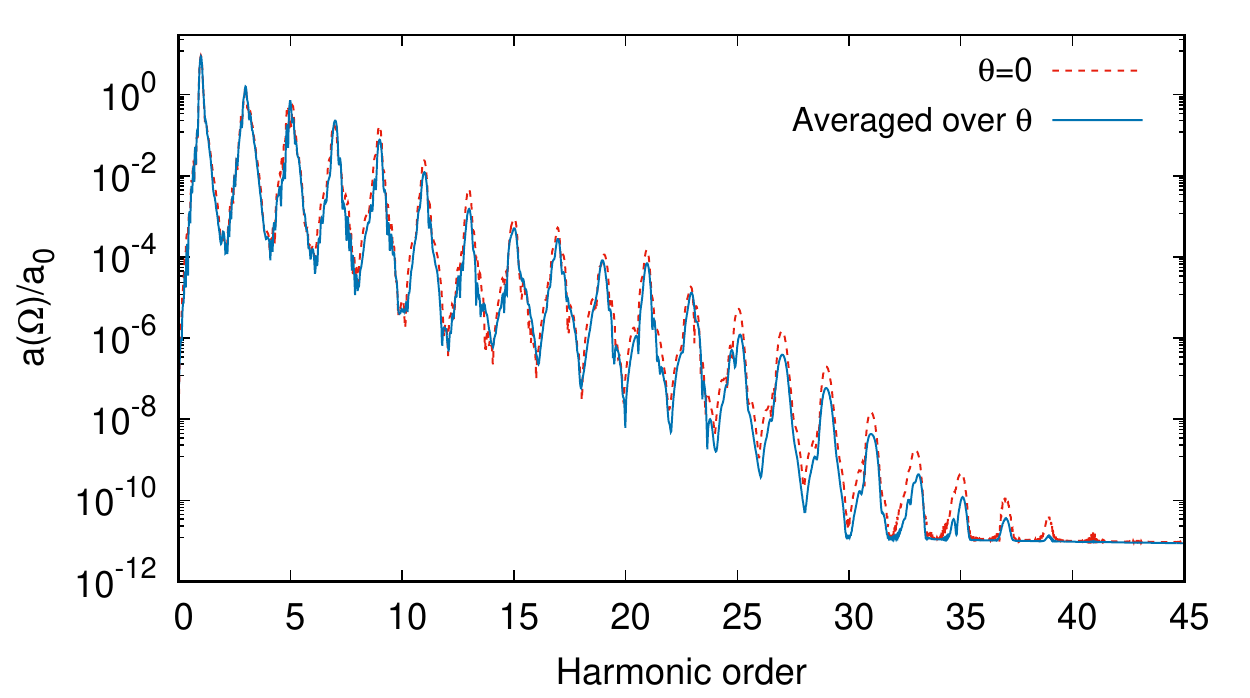}
\caption{The averaged HHG spectrum for GQD$_{54}$. The averaging is
performed over hexagon rotation angle with respect to wave polarization
direction. The wave amplitude is taken to be $E_{0}=0.5\ \mathrm{V/\mathring{%
A}}$.}
\end{figure}

In the context of the experimental realization of HHG in GQDs, it's crucial
to account for the arrangement of GQDs randomly distributed across the 2D
surface. The angle, denoted as $\theta $, between one side of the hexagon
(see Fig. 1) and the polarization vector of the incident wave plays a main
role in determining the resulting HHG spectra. Therefore, it becomes
necessary to consider a range of randomly distributed $\theta $\ values and
subsequently perform averaging over these angles. Taking into consideration
the symmetry inherent to the GQDs, we performed calculations with $100$\
randomly distributed $\theta $\ values spanning the range from $0$\ to $\pi
/3$\ and then averaged the results. The results of this averaged HHG
spectrum, in addition to the spectrum obtained at $\theta =0$, are presented
in Figure 15. Note that for a single GQD when $\theta \neq 0$\ there is also
HHG in the perpendicular to laser polarization direction, which is averaged
to zero. As depicted, the averaged HHG spectrum exhibits a reduction when
compared to the optimal value at $\theta =0$. Nonetheless, thanks to the high 
symmetry inherent in the considered GQD, this reduction does not have a critical impact 
on the overall result.

\section{Conclusion}

We have presented a comprehensive investigation into the extreme nonlinear
optical response of the hexagonal graphene-based quantum dots. Specifically,
we focused on GQDs composed of $24$ and $54$ carbon atoms, which represent
inversion symmetric configurations commonly found in these systems. Our
study employed an accurate quantal calculation of the HHG spectra using a
mean-field approach that accounts for many-body Coulomb interaction. By
solving the evolutionary equations for the single-particle density matrix,
we disclosed intricate fine structures within the HHG spectra and observed a
significant enhancement in mid-plateau and near-cutoff harmonics, which can
be attributed to the effect of the long-range correlations. Our findings
highlight the crucial role of the Coulomb interaction in determining of the
harmonics intensities and the cutoff position. To gain deeper insights into
the high-harmonic generation mechanisms across the different energy ranges,
we employed a detailed wavelet time-frequency analysis. Such analysis
uncovered intricate spectral and temporal fine structures, shedding new
light on the underlying processes involved. Additionally, our investigation
revealed a strong dependence of the HHG spectra on the number of particles,
indicating a preference for GQDs with a larger number of particles. While
our results are presented for hexagonal GQDs, it is reasonable to expect
similar outcomes for experimentally accessible GQDs with triangular and
rectangular shapes, since the obtained results do not rely on the shape of
the GQD. This broadens the applicability of our findings. Overall, our study
provides a basic insight into the nonlinear optical response of GQDs and
contributes to a better understanding of characteristics of such a
significant phenomenon as HHG. However, it's important to note that our
approach relied on TB theory, which includes several free parameters
adjusted with respect to infinite graphene sheet. While this TB-based
approach has provided valuable insights into GQD dynamics and their response
to laser fields, we recognize that for a more detailed and precise
description of the system's behavior, a more advanced method, such as
time-dependent density functional theory (TDDFT), is required. In our future
work, we plan to perform calculations within the framework of TDDFT. This
approach will allow us to account for higher-order correlations without the
need for adjustable parameters, leading to a more comprehensive
understanding of GQD dynamics and their nonlinear optical response.

\begin{acknowledgments}
The work was supported by the Science Committee of Republic of
Armenia, project No.21AG-1C014.
\end{acknowledgments}

\bibliographystyle{apsrev4-1}
\bibliography{bibliography}

\end{document}